# Should You Take Investment Advice From WallStreetBets? A Data-Driven Approach

*May 2021*

**Tolga Buz**
Hasso Plattner Institute, Potsdam
tolga.buz@hpi.de

**Gerard de Melo**
Hasso Plattner Institute, Potsdam
gdm@demelo.org

## Abstract

*Reddit's WallStreetBets (WSB) community has come to prominence in light of its notable role in affecting the stock prices of what are now referred to as* meme *stocks. Yet very little is known about the reliability of the highly speculative investment advice disseminated on WSB. This paper analyses WSB data spanning from January 2019 to April 2021 in order to assess how successful an investment strategy relying on the community's recommendations could have been. We detect buy and sell advice and identify the community's most popular stocks, based on which we define a WSB portfolio. Our evaluation shows that this portfolio has grown approx. 200% over the last three years and approx. 480% over the last year, significantly outperforming the S&P500. The average short-term accuracy of buy and sell signals, in contrast, is not found to be significantly better than randomly or equally distributed buy decisions within the same time frame. However, we present a technique for estimating whether posts are proactive as opposed to reactive and show that by focusing on a subset of more promising buy signals, a trader could have made investments yielding higher returns than the broader market or the strategy of trusting all posted buy signals. Lastly, the analysis is also conducted specifically for the period before 2021 in order to factor out the effects of the GameStop hype of January 2021 – the results confirm the conclusions and suggest that the 2021 hype merely amplified pre-existing characteristics.*

## Introduction

Members of Reddit's r/WallStreetBets (WSB) community often make a point to emphasize that their posts should not be deemed as constituting financial advice. Yet, the forum appears to have profoundly affected a number of specific stock prices, most prominently in the January 2021 GameStop short squeeze. Some of these have since been known as *meme stocks*, reflecting the social contagion arising from trending posts in the forum. While WSB is widely known for its highly speculative and unorthodox forms of sharing and debating investment strategies, little is known about the reliability of the disseminated recommendations. This paper presents an analysis based on a dataset comprising WSB posts spanning from January 2019 to April 2021 in an attempt to shed light on this question. The principal goals are to identify the portfolio of most discussed stocks and assess how reliable the community's buy and sell signals would have been if taken as investment advice. As the analysis shows that not all investment decisions shared on WallStreetBets lead to financial success, we further develop means to distinguish different sorts of posts and identify proactive buy signals with promising prospects.

Since the stock market is a complex system affected by a multitude of diverse stakeholders, events, and influences, we investigate to what extent specific patterns in the posts on r/WallStreetBets may have anticipated corresponding stock price movements that occurred later. We also compare analysis results based on data from the *pre-hype* and *post-hype* time windows, as the rapid rise to prominence of the WSB community in January 2021 also entailed an explosive growth in user numbers and generated content.

The following sections provide background information on r/WallStreetBets and on relevant research related to this matter as well as describe the dataset and how it has been collected and analysed. The subsequent sections introduce our techniques to identify the most popular stocks within thousands of text snippets and the characteristics these stocks share, followed by an analysis of how reliable the community's discussions and buy/sell signals would have been if taken as investment advice at the time of posting. Within each of the sections, the adopted methodology is explained, along with analysis results as well as specific examples of characteristic stocks.

## Background

Reddit is a social media platform that provides a place for communities (known as *subreddits*) focusing on specific topics, such as humorous memes, politics, relationship advice, particular sports teams, or computer games, among numerous others. Reddit was established in 2005 by Steve Huffman and Alexis Ohanian, and has since grown to host over 100,000 communities with more than 52 million daily active participants, accumulating more than 50 billion monthly views (Reddit, Inc. 2021). Members can post submissions (e.g., texts, links, images, videos), comment on them, use the up- or downvote buttons to rate submissions and comments, and even gift authors with awards purchased beforehand.

The subreddit r/WallStreetBets was created on January 31, 2012 at http://www.reddit.com/r/wallstreetbets. On January 1, 2019 the number of subscribers was approximately 450,000, while on January 1st, 2021, this figure had grown to approximately 1,760,000. At the time of writing, the subreddit has approximately 9.6 million subscribers after the community's popularity increased rapidly from 1.7 million to approximately 8.5 million within January 2021 (Subredditstats.com 2021). While the website subredditstats.com places r/WallStreetBets at rank 50 out of all subreddits regarding the subscriber count at the time of writing, the same source places the community among the top-ranked with respect to the amount of comments and posts per day (ranks 2 and 3, respectively – at the time of writing).

The subreddit describes itself as "a community for making money and being amused while doing it. Or, realistically, a place to come and upvote memes when your portfolio is down." Like many other Reddit communities, WSB is notorious for its unique slang and the pervasive use of offensive terms. For instance, members of the community are officially referred to as "degenerates".

In January 2021, WSB saw an unprecedented rise in popularity and news coverage after the community focused discussions on a series of stocks that were in part deemed undervalued while simultaneously exhibiting a high short interest ratio, i.e., the ratio of shares being sold short (or *shorted*) by financial institutions. Short-selling a stock refers to the practice of borrowing shares of a stock in order to sell them immediately with the goal of buying them back later at a lower price. This practice is a speculative move often employed on stocks that are considered overvalued and expected to lose value in the near future.

While discussions of potentially undervalued investment opportunities with growth potential are common in investment-focused online communities, the fact that the GameStop stock showed a pronounced short interest by institutional investors led to the situation being portrayed as an ideological David-and-Goliath battle of small retail investors rallying against hedge funds: by buying and holding the stock, its price could be driven up, thus forcing the financial institutions to close their short positions at a significant loss, which in turn drove the stock price up even further, a phenomenon known as a *short squeeze*, at which point the retail investors could sell their shares at a large profit.

This made r/WallStreetBets a place where the broader community of Reddit users could come together to unite in a movement, driven by the prospect of large financial gains through risky investments, with the added appeal of supposedly advancing the greater cause of punishing the financial institutions, particularly hedge funds. The latter are accused of ruining many people's lives during the financial mortgage crisis of 2008, among other events.

# Related Work

Social media has been considered a valuable source of data for detecting public sentiment on topics such as politics, companies, brands, and products, and is frequently studied especially in behavioural finance and decision sciences. Research suggests that there are important ties between social media sentiment and macroeconomic factors such as consumer confidence (Daas and Puts 2014) and that the general social media sentiment may affect a firm's stock performance (Yu et al. 2013). Extensive research has attempted to show how particular cues from social media can enable predicting stock price changes (Duz Tan and Tas 2021; Nguyen et al. 2015; Sul et al. 2017).

Another line of work focuses on the forms of interaction in modern online platforms. For instance, it has been shown that emotionally charged messages in social media can spread information quickly (Stieglitz and Dang-Xuan 2013). Messages spread on social media, in some circumstances driven by bots, have been known to empower social movements (Manikonda et al. 2018) and to have political influence at an international scale (Gorodnichenko et al. 2018; Howard et al. 2011). A recently published paper based on research in 2020 investigated the social interaction on r/WallStreetBets (Boylston et al. 2021), explaining the nature of the community's conversations and language.

Fuelled by the 2021 GameStop hype on r/WallStreetBets, recent research has provided a number of insights on the dynamics and background of the matter. This includes studies on the social dynamics within the WSB community that led to the hype (Lyócsa et al. 2021; Semenova and Winkler 2021) and on the idea of retail investors fighting against Wall Street (Chohan 2021; Di Muzio 2021). Other studies focused on the financial mechanisms driving the sudden price spike (Aharon et al. 2021; Feinstein 2021), and on the effect of retail traders on prices and volatility, along with their participation in transactions (Eaton et al. 2021; Hasso et al. 2021; van der Beck and Jaunin 2021). Further research considered the implications of the events for market regulators and brokerages (Jones et al. 2021; Macey 2021; Umar et al. 2021).

The variety of related research shows that r/WallStreetBets and the GameStop hype can be investigated from multiple angles. However, most recent research focuses on socio-economic and general market effects and implications. Little is known about the financial skills of WSB traders and the merits of their investment advice (Bradley et al. 2021). Additionally, there is a lack of longitudinal research from an information science perspective – studying data that span a longer time frame and a large number of different postings and assessing longer-term effects and trends. This paper presents new insights from this data-driven perspective on the WSB community's proficiency in order to evaluate how well investors following the submissions and buy recommendations in the forum would have performed financially since January 2019.

# Dataset Compilation

In order to assess the reliability of advice shared on WSB over a long period of time, we compiled a dataset and in the following provide an overview of the its size and characteristics.

## Data Acquisition

The dataset was collected with a script accessing the Pushshift Reddit archiving service (Baumgartner et al. 2020) via its web API. The latter allows requesting data exports in JSON format based on submissions or comments. In order to compile a representative dataset, all submissions with an upvote score greater or equal 1 from January 1, 2019 to April 4, 2021 were retrieved in a first step. This constitutes a sufficiently long time span to analyse the activity on WSB before and after the COVID-19 pandemic affected stock markets in March/April 2020 as well as before and after the January 2021 GameStop incident. The upvote score threshold serves to reduce noise, as there are many posts with an upvote ratio below 50% (i.e., more downvotes than upvotes, leading to a score of zero) – these submissions often do not meet quality or community standards, or address topics or opinions that are not accepted in the community. Additionally, all submissions deleted by their authors or WSB moderators were eliminated from the dataset, as a deletion is generally due to a violation of community guidelines or due to a submission's unpopularity.

Due to the Pushshift API's request size limitation, our script starts soliciting the most recent submissions and then repeatedly issues new requests by iteratively moving the `before` parameter further back so as to receive results preceding the previous iteration's earliest submission, until the stop condition is met, i.e., the desired time period has been processed. The results include additional metadata, e.g., the time of posting, the author identifier, a potential category tag ("flair"), etc. While the service also offers the option of collecting comments provided in response to submissions, in this analysis we focus on submissions only, because they represent the main topics that a user encounters when visiting and browsing r/WallStreetBets. Comments are typically short replies to the topic presented in the submission, but are not as prominently displayed and not as carefully curated as submissions. They are very heterogeneous in terms of their information value. If a comment provides substantial criticism or corrections, the authors of an original post may incorporate such information into the post through subsequent edits.

|  | Titles | | Submissions | |
|--|--------|--------|-------------|--------|
|  | incl. s/w | excl. s/w | incl. s/w | excl. s/w |
| Words | 7,671,264 | 5,221,098 | 26,414,494 | 16,214,340 |
| Characters | 41,967,515 | 31,099,220 | 159,353,612 | 111,364,356 |
| Avg. Text Length | 9.86 | 6.71 | 33.94 | 20.83 |
| Vocabulary Size | 192,659 | | 520,022 | |
| # Texts (2019–2021) | 778,288 | | 238,001 | |
| # Texts (2021 only) | 494,885 | | 124,704 | |

**Table 1. Dataset and corpus statistics (s/w = stop words)**

It should be noted that the Reddit data provided by the Pushshift service may deviate from the original data in minor ways. Most importantly, for some posts, the provided upvote scores are incomplete and thus substantially lower than the true value. Additionally, in the beginning of February 2021, the service was offline for multiple days. We acknowledge the risk that due to similar outages in the past, Pushshift may not have been able to scrape Reddit consistently. However, other relevant metadata, e.g., creation date and number of comments, were always correct in our sample comparison. Thus, we conclude that apart from the score value, the data is sufficiently reliable to conduct the analysis. Overall, the Pushshift service provides a valuable source of historic posts that may otherwise by non-trivial to collect.

### *Dataset Statistics*

The full dataset consists of 778,288 submissions posted between January 1, 2019 and April 4, 2021. We can distinguish two kinds of content:

1. Titles: Texts from all collected submission's titles
2. Submissions: Texts from all collected submissions containing a text body ("selftext"), while disregarding posts only consisting of a title plus an image or video.

Table 1 provides an overview of word and character counts for titles and submissions, reporting figures both including and excluding a pre-defined list of stop words.[1] The vocabulary sizes are based on the cleaned corpora, excluding stop words and common punctuation, and after lowercase character normalization. A significant difference in vocabulary sizes is observed, as titles and submissions by design have substantially different lengths. In general, the vocabulary sizes are fairly large, which in part stems from the fact that the corpus consists of user-generated text from social media and hence also harbours substantial numbers of typos, emojis, symbols, and similar phenomena that can increase the vocabulary size.

The overall dataset consists of 778,288 submissions, of which 238,001 include a body text. A significant part of these submissions was posted in 2021 (494,885 in total, 124,704 with body text), due to the aforementioned dramatic expansion of the community's subscriber count and amount of posted content. Approximately 64% of all submissions (titles) and 53% of submissions with body text (submissions) have a creation date in 2021, while the rest of the dataset spans across 2019 and 2020. This shows how much traction the subreddit has gained in 2021, in large part due to the GameStop hype, and is in line with the growth of subscribers from 1.7M to 8.5M within January 2021.

---

[1]Stop word list accessible via: https://github.com/explosion/spaCy/blob/master/spacy/lang/en/stop_words.py

On WSB, a system of tags, known as link *flairs*[2], serve as labels to categorize posts as *Discussion*, *DD* (due diligence), *News* (concerning specific stocks or markets), *Gain* and *Loss* (recounting gains or losses that need to surpass a certain monetary value threshold in order to be approved), or *YOLO* (high-risk high-value investments), among others. In our dataset, 718,517 submissions are labelled with such a tag either by their author or by the moderators. A review of this feature shows that the most common submission tags are distributed as follows: 23.5% *Discussion*, 22.1% *Meme*, 12.6% *YOLO*, 8.6% *News*, 7.7% *Gain*, 7.6% *Shitpost*, 4.9% *Loss*, 4.5% *DD*, with the rest distributed across other tags. This suggests that members of WSB use it to discuss ideas and opinions and share news articles, while posting twice as much about their gains compared to their losses, but even more about YOLO endeavours, i.e., high-risk investments. Additionally, WSB as well appears to serve as a source of entertainment, regarding the large ratio of memes and "shitposts", although one could argue that *YOLO*, *Gain*, and *Loss* reports as well may provide entertainment value. Among the top 100 most successful individual posts in terms of their upvote score at the time of writing, the most prominent categories of tags are *News* and *Meme* (32 each), followed by *Discussion* (9), *Loss* (8), *YOLO* (7), *Chart* (6), *Gain* (3), *Satire* (2), and *DD* (1).

An analysis of the creation timestamps of submissions yields several insights on WSB post activity and suggests that a sizeable portion of users reside in the US and visit the community during daytime and office hours:

- On weekdays, the highest post activity is observed during 2 p.m. – 6 p.m. UTC, which translates to 10 a.m. – 2 p.m. EST (US East Coast) and 7 a.m. – 11 a.m. PST (US West Coast).
- The activity peak on weekdays is achieved at 4 p.m. UTC / 12 p.m. EST / 9 a.m. PST.
- On weekends, the peak activity shifts to later hours: between 12 a.m. – 3 a.m. UTC / 8 p.m. – 11 p.m. EST / 5 p.m. – 8 p.m. PST.

Before 2021 and thus before the rapid growth in users, activity during weekdays peaked at 10 p.m. UTC / 6 p.m. EST / 3 p.m. PST, closely followed by a secondary peak at 4 p.m. UTC. This seems to have changed in 2021 as the main peak shifted to 4 p.m. UTC, which corresponds to noon (East Coast) or the start of the work day (West Coast). Hence, with the sudden growth of the community, WSB's activity has increased especially during working hours, which could be due to authors spending more time in the forum during work, possibly because they work in a related field or are distracted from work and prefer to participate in the WSB community instead.

## The WallStreetBets Portfolio

This section reviews the stocks and ETF products discussed on r/WallStreetBets and comes to the conclusion that the overall portfolio, i.e., the set of stocks most frequently discussed by the community, has performed better than the broader market (in this case identified by the SPDR S&P500 ETF, which is also one of the most discussed tickers) in the recent past. Specifically, we consider the 1-year and 3-year time frame.

### *Methodology: Detecting Stock Tickers*

Each stock can be identified by its ticker, which is usually an abbreviation related to the company's name or main product (e.g., AAPL for Apple Inc., CRM for Salesforce). These tickers are conventionally written in capital letters and sometimes adorned with a preceding "$" (e.g., $AAPL, $CRM) in order to identify them as stock tickers – this is particularly helpful when it could otherwise be mistaken for a regular word or abbreviation (e.g., $COST for Costco Wholesale Corporation, $SHOP for Shopify Inc., $NOW for ServiceNow Inc., $CAR for Avis Budget Group Inc.).

Unfortunately, user-generated data from online communities tends to require cleaning and initial analysis has shown that the posters frequently omit the preceding dollar sign, in many cases even using lowercase let-ters for the tickers or writing entire sentences or posts in uppercase letters (e.g., "BUY THIS STOCK NOW"). Additionally, many commonly used abbreviations of business- or sector-related terms can be mistaken for stock tickers as well, e.g., GM for General Manager or General Motors Co., GDP for a country's Gross Domestic Product or Goodrich Petroleum Corp, CAC for Customer Acquisition Cost or Camden National Corp. There is also the possibility that posts about a particular stock may include mentions of additional companies, e.g., if the bank JPMorgan Chase & Co. (JPM) has commented on buying or selling a specific stock like GameStop. This makes it challenging to detect stock tickers correctly without counting other words as false positives or alternatively risking that important stock tickers are missed by filtering out too many words.

---

[2]Explained in the guidelines accessible via https://www.reddit.com/r/wallstreetbets/wiki/linkflair

For our study, in order to identify stock tickers in the corpora, the texts are tokenized as sequences of words while removing punctuation that commonly appears adjacent to words (e.g., commas, periods, parentheses), then filtered for short tokens with 2-5 uppercase alphabetic characters or tokens starting with "$" followed by 1-5 alphabetic characters. Single characters appearing without a "$" are excluded, because initial analyses showed that most of these are false positives and single-character stock tickers like Ford Motor Company ($F) are usually mentioned with the dollar sign, while occurrences without "$" usually refer to a popular swear word starting with "F". Tokens without an affixed dollar sign are counted if they are included in a combined list of known stock[3] and ETF[4] tickers, while not part of a list of custom stop words[5]. Tokens starting with a "$" are counted if followed by uppercase letters, which excludes monetary values such as $1000, but allows the inclusion of tickers not included in the predefined stock ticker list. In case of stock tickers that could be used as normal words or are single letters, the WSB community frequently uses the "$" to specify a stock ticker, so this identification strategy works well in practice.

While a high-precision method to detect stock tickers could be to focus on tickers mentioned with a preceding dollar sign only, this would have very low recall, ignoring a substantial proportion of all stock mentions. An evaluation of the dataset reveals that a much larger ratio of stock symbols is mentioned without a "$". For instance, only 13.4% of all $GME (GameStop) mentions and 11.6% of all $MSFT (Microsoft) mentions had the preceding "$". The strategy proposed above instead constitutes a semi-automatic method of detecting the vast majority of stock tickers, while reducing the amount of false positives caused by mistaking commonly used words as stock mentions.

## Analysis

Invoking the identification technique described above, the 100 most frequently mentioned stock and ETF tickers can be collected for different time windows: the dataset's full range (from January 2019 to April 2021), as well as each year separately (2019, 2020, 2021).

Grouping the tickers contained in these lists by sector shows that the WSB community focuses primarily on stocks from the sectors *Consumer Cyclical* (non-essential consumer goods, e.g., apparel, products related to leisure activities, automobiles), *Technology* (manufacturers of electronics, software, semiconductors, etc.), *Healthcare* (including pharmaceuticals, cannabis companies), and *Communication Services* (including interactive media, entertainment, telecommunications), as reported in Table 2.[6] The distribution of the sectors' market capitalization shows that WSB does not discuss an identically distributed set of sectors, but instead appears to focus on those that consumers are most able to relate to due to regular exposure to their products or services.

Particular attention is given to stocks of comparably young companies, which are often considered growth stocks – these tend to be riskier due to higher volatility, but exhibit a higher potential to increase in value (in contrast to low-volatility value stocks of dividend-paying, older companies such as The Coca-Cola Company): The only three sectors (*Technology*, *Consumer Cyclical*, and *Communication Services*) that have performed better regarding their 3-year growth than the S&P500 index, which is considered a good reference for the broader market, are all ranked within the community's top four sectors. Regarding the 1-year growth, WSB's focus on the same three sectors has been successful as well, but the community seems to have missed out on the stronger growth in the *Basic Materials* and *Energy* sectors. Additionally, *Industrials* and *Financial services* have received some attention from WSB (consistently in the top 5–6) and grew above average during the last year. However, there has also been a focus on the healthcare industry, which has performed worse than the broader market over the 3-year and 1-year time windows. The discussed healthcare stocks are mainly drug manufacturers, especially companies selling cannabis or COVID-19 vaccines.

---

[3]This was exported from the NASDAQ Stock Screener (Nasdaq, Inc. 2021).
[4]This was extracted manually from the WSB community's discussions about ETFs.
[5]A manually created list of short tokens that are often used as words or abbreviations and could be mistaken as stock symbols due to their frequent occurrence on WSB in uppercase.
[6]The columns each add up to slightly less than 100 (more precisely: 93–95), because a sector categorization could not be obtained for all stock tickers.

| | Count of top WSB tickers | | | | Financial data | | |
|---|---|---|---|---|---|---|---|
| Sector | 2019–2021 | 2021 | 2020 | 2019 | Market Cap. | 1-Yr Change | 3-Yr Change |
| Consumer Cyclical | 21 | 22 | 24 | 20 | $8.90T | +87.82% | +58.46% |
| Technology | 20 | 17 | 22 | 24 | $13.30T | +75.70% | +86.57% |
| Healthcare | 14 | 14 | 11 | 13 | $7.39T | +37.35% | +35.10% |
| Comm. Services | 13 | 9 | 10 | 16 | $6.24T | +68.74% | +54.44% |
| Industrials | 10 | 11 | 11 | 6 | $5.45T | +79.70% | +26.26% |
| Financial Services | 7 | 10 | 5 | 6 | $8.13T | +79.26% | +17.98% |
| Consumer Defensive | 3 | 1 | 5 | 6 | $4.25T | +19.03% | +23.40% |
| Basic Materials | 3 | 5 | 2 | 0 | $2.59T | +89.46% | +31.69% |
| Energy | 1 | 1 | 2 | 2 | $2.59T | +105.69% | -25.67% |
| Utilities | 1 | 0 | 2 | 2 | $1.54T | +13.25% | +22.97% |
| Real Estate | 1 | 4 | 0 | 0 | $1.50T | +33.11% | +23.89% |
| S&P500 (for reference) | | | | | | +63.18% | +42.19% |

**Table 2. Distribution of sectors for top 100 mentioned stocks per time window with market capitalization, 1-year, and 3-year change (as of March 20, 2021) per sector**

On the whole, we can conclude that WallStreetBets' focus on stocks from consumer- and technology-related sectors correspond to those that have generally seen beneficial outcomes in terms of long-term sector growth compared to the broader market performance. This conclusion is supported when taking a closer look at the most popular stocks and their products. 21 tickers have consistently been part of the 100 most frequently discussed stock tickers for each of the three years (including $SPY, which will be considered a reference value for the broader market). These will be referred to as the "WSB portfolio" in the following and can be grouped based on the respective industry:

- **Entertainment & Internet:** AMC Entertainment Holdings Inc., Facebook Inc.
- **Auto Manufacturers & Retail:** General Motors Company, NIO Inc., Tesla Inc., Amazon.com Inc., Alibaba Group Holding Limited, GameStop Corporation,
- **Consumer Technology & Semiconductors:** Apple Inc., Advanced Micro Devices Inc., Intel Corporation, NVIDIA Corporation
- **Technology & Software Infrastructure:** BlackBerry Limited, Nokia Corporation Sponsored, Microsoft Corporation
- **Cannabis Companies:** Aurora Cannabis Inc., Tilray Inc.
- **Industrials:** General Electric Company, Boeing Company
- **ETFs:** SPDR S&P 500, Invesco QQQ Trust

A review of this portfolio (as given in Table 3) reveals that 10 out of 18 tickers have performed better than the S&P500 ETF ($SPY) with respect to the 3-year change (2 additional ones, NIO and TLRY, went public at a later time), and 14 out of 20 have outperformed the S&P500 ETF with respect to the 1-year change – both as of March 20, 2021. While $SPY has achieved 51.93% growth over three years and 72.96% over the last year, a hypothetical, equally-weighted portfolio of these most frequently and consistently discussed stocks from WSB has grown by 198.60% (mean) / 72.53% (median) over the last three years and by 483.20% (mean) / 103.74% (median) over the last year. Thus, the WSB portfolio has grown faster than the broader market on average – particularly following the market crash caused by the COVID-19 pandemic. In conclusion, this means that on the longer term of two to three years as well as throughout the last year, investing in the most popular stocks on WSB most likely would have been a good investment decision (compared to overall market growth).[7]

| Ticker | Company name | 1-Year Change (%) | 3-Year Change (%) |
|--------|--------------|------------------:|------------------:|
| AAPL | Apple Inc. | 110.99 | 183.88 |
| ACB | Aurora Cannabis Inc. | 15.75 | -89.29 |
| AMC | AMC Entertainment Holdings, Inc | 336.68 | 20.10 |
| AMD | Advanced Micro Devices, Inc. | 99.60 | 611.03 |
| AMZN | Amazon.com, Inc. | 66.57 | 93.82 |
| BA | Boeing Company (The) | 169.26 | -20.88 |
| BABA | Alibaba Group Holding Limited | 32.26 | 20.53 |
| BB | BlackBerry Limited | 223.96 | -16.28 |
| FB | Facebook, Inc. | 93.76 | 72.53 |
| GE | General Electric Company | 103.74 | 3.82 |
| GM | General Motors Company | 229.77 | 76.24 |
| GME | GameStop Corporation | 5226.33 | 1446.41 |
| INTC | Intel Corporation | 42.74 | 33.10 |
| MSFT | Microsoft Corporation | 69.41 | 157.43 |
| NIO | NIO Inc. | 1706.25 | N/A |
| NOK | Nokia Corporation Sponsored | 51.13 | -25.67 |
| NVDA | NVIDIA Corporation | 150.00 | 107.47 |
| QQQ | Invesco QQQ Trust, Series 1 | 84.49 | 90.68 |
| **SPY** | **SPDR S&P 500 ETF** | **72.96** | **51.93** |
| TLRY | Tilray, Inc. | 594.52 | N/A |
| TSLA | Tesla, Inc. | 665.88 | 954.37 |
| Mean | | 483.20 | 198.60 |
| Median | | 103.74 | 72.53 |

**Table 3. Consistently and frequently discussed stock tickers with 1-year and 3-year price development (as of March 20, 2021)**

## Investment Advice Reliability

The previous section has shown that despite an emphasis on a few specific sectors, WSB's portfolio of most consistently discussed stocks has performed better than the broader market on average over the last three years (as of March 20, 2021), based on an analysis of the longer-term success of r/WallStreetBets' most popular sectors. This section, in contrast, focuses on evaluating the short- to medium-term success of potential investment decisions, identifying spikes in the community's activity and comparing them to movements on the stock market.

We do not assume that all stock price movements are reflected in the r/WallStreetBets community's discussions, let alone caused by the community. Instead the stock market is considered to be a complex system affected by countless different events and stakeholders, of which one might be the WSB community. Therefore, this section focuses on comparing signals and activity on r/WallStreetBets to movements in the stock market in order to assess how activity on WSB can be interpreted or utilized, possibly to anticipate market movement. We investigate whether specific types or intensity levels of WSB activity indicate a good timing for an investment decision followed by a positive price development after buy signals, and to what extent specific investment advice from WSB has been successful in the recent past.

Success in financial markets is a matter of definition and time – in the following, we focus on comparing daily

---

[7]Naturally, the timing of an investment can have a significant impact. It must be noted that the 1-year change coincides with the low-point of stock prices right after the steep decline due to the COVID-19 outbreak in February/March 2020, which explains why many 1-year performance gains are higher than the respective 3-year values.

WSB activity with stock price changes at specific points in time relative to the date in question, e.g., after one day, after three days, after a week, etc. We hence evaluate how the value of the investment developed over specific short-term time frames, after a hypothetical investor decides to follow a WSB buy signal for a specific stock on a specific day. If the price is higher after a specific time window has passed, e.g., one day, the buy signal is rated as successful for this time window.

## Methodology

In order to measure stock-related activity on r/WallStreetBets, multiple features are derived from the underlying data. While the simplest approach is just counting mentions of a stock ticker or name, a more targeted assessment can be conducted by counting transaction-related words that occur in the context of the ticker mentions, specifically "buy", "hold", "sell", "call", "put" and similar words, that are mentioned in the same submission as the respective stock ticker. The collected word and ticker counts for each submission are aggregated per day in order to create a daily summary of activity related to each stock ticker. This daily summary is then joined with financial data related to each stock from the Yahoo! Finance API – for every day, the stock's closing price, daily high and low, and trading volume are added. As the stock markets are closed on weekends and public holidays, the volume on those days are set as zero and the previous day's closing price is taken as that day's closing price as well as daily high and low.[8] Based on the relative difference between high and low values with regard to the closing price, a relative volatility value is calculated (which is equal to zero on closed market non-trading days). This results in a dataset consisting of mention counts as well as financial data at a daily granularity level for each selected stock.

To enable a more detailed analysis, further features have been added: For each day, the word counts of "buy" and "sell" are compared – whichever count is higher defines if the day is regarded as a *buy signal* or *sell signal*, which is treated as a piece of investment advice in this context. Additionally, for every day, we compute relative changes of the closing price per respective date compared to 1 day, 3 days, and 1 week before, as well as 1 day, 3 days, 1 week, 1 month, and 3 months later. Thus, for a past date like January 1, 2019, one can easily look up by how many percentage points the price will have changed one month later. Based on these price differences, various boolean features have been added to identify the following cases (where $x$ days refers to 1 day, 3 days, 1 week; $y$ days refers to 1 day, 3 days, 1 week, 1 month, 3 months):

- Price increased or decreased in preceding $x$ days
- Price increased or decreased in subsequent $y$ days
- Price dip (decrease in preceding $x$ days and increase $x$ days later)
- Buy signal after price decreased in preceding $x$ days
- Reactive buy signal (price increase in preceding $x$ days is higher than price change $x$ days later)
- Proactive buy signal (price increase after $x$ days is higher than price change in preceding $x$ days)
- Buy accuracy (defined as true if price increased within $y$ days after buy signal, false if decreased)
- Sell accuracy (defined as true if price decreased within $y$ after sell signal, false if increased)

These features are invoked to assess the success rate of WSB submissions' investment recommendations: For evaluation of the success rate of buy and sell signals, all signals per selected stock ticker are tracked with the respective price changes before and after the day they occurred in order to decide if a hypothetical investment made on a specific day with a buy signal would have gained or lost value over specific time windows. A buy signal is considered successful within a specific time window if and only if the closing stock price increased within the same window – if it decreases, the buy signal is considered a failure. Accordingly, sell signals are evaluated for the respective opposite cases. Each time window is reviewed separately (one day, three days, one week, one month, and three months).

Our study produced such daily data summaries for each of the WSB portfolio tickers mentioned in the previous section as well as for the overall S&P500 index from Yahoo Finance ($^GSPC) as an additional reference. This results in a dataset containing daily values about WSB activity, stock price, volatility, and volume for

---

[8]While trading is not possible on weekends or holidays, this method helps in assessing price movements that are longer than three days – if someone owned a specific stock on a Sunday at a price of $30 (closing price of the previous Friday) and a month later on Sunday the price was at $40, this can be considered a positive price development.

each of the portfolio's stocks.

### Short-Term Performance

On average, the WSB portfolio stocks have grown 3–4 times more than the S&P500 – if an investor bought shares of the WSB portfolio on every day between January 2019 and December 2020, the investments would have grown by more than 16% on average after 3 months, while the S&P500 would have grown slightly less than 5% on average in the same time frame (see Table 4). Furthermore, the table shows that if the same investor had chosen to buy on days of WSB activity regarding the respective portfolio tickers, the average 3-month growth would have reached 27.32% (buying if the ticker was mentioned), 32.03% (if additionally a buy signal was detected), 28.53% (if additionally a sell signal was detected). A similar pattern can be observed for the shorter time frames of a day, a week, and a month. This shows that when stocks were bought on days of activity on WSB, the average price increase of the portfolio tickers would have been higher than when bought over all days, performing better than the broader market with an even larger difference. Overall, this indicates that on average the WSB portfolio outperformed the S&P500 in short-term investment windows since 2019. Nonetheless, it should be noted that sell signals have been quite unsuccessful within the reviewed time windows – in practice interpreting sell signals as buy signals would have generated a similarly beneficial outcome as relying on actual buy signals, with even higher average price increases for short term windows of one day and one week.

|  |  | S&P500 | WSB Portfolio when bought on specific days: | | | |
|  |  |  | All | w/ Mention | w/ Buy Sig. | w/ Sell Sig. |
|---|---|---|---|---|---|---|
| Avg. price change (%) after | 1 day | 0.07 | 0.17 | 0.29 | 0.29 | 0.65 |
|  | 1 week | 0.44 | 1.24 | 2.07 | 2.19 | 3.77 |
|  | 1 month | 1.79 | 5.14 | 7.77 | 9.88 | 8.12 |
|  | 3 months | 4.89 | 16.65 | 27.32 | 32.03 | 28.53 |
| Average | mentions | - | 28.28 | 51.69 | 89.50 | 15.53 |
|  | daily volat. | 0.8% | 2.9% | 3.7% | 4.2% | 3.7% |
|  | daily volume | 2.99B | 25.56M | 32.09M | 37.95M | 32.59M |

**Table 4. Average price change per time window, mentions, daily volatility, and daily volume – for S&P500 and WSB portfolio (calculated using all days or only days with ticker mentions or buy/sell signals, January 2019 – April 2021)**

The average volatility of the S&P500 is low, as the index value is an average of 500 stocks. The WSB portfolio stock's average daily volatility increases when there is activity on WSB, especially if there are buy signals. The WSB portfolio's 21 stocks unsurprisingly generate a much lower average trading volume than the S&P500, which, as the name implies, consists of 500 stocks. However, with the simplifying assumption that an average stock listed in the S&P500 corresponds to a daily trading volume of approx. 5.98 million USD (1/500 of the S&P500 volume), one finds that the average WSB portfolio stock is traded with a five times higher volume compared to the average S&P500 stock. The average daily trading volume is significantly higher on days with WSB activity, especially if a buy signal is detected (approx. 49% higher than on average). The values imply that WSB activity increases when markets are volatile and volume is high or possibly that increased WSB activity may lead to higher volatility and volume in financial markets.

However, it is difficult to prove to what extent WSB's activity may have a non-negligible influence on particular stock prices, volatility, or volume figures – as the stock market is subject to many different external factors. Nonetheless, one may consider all significant WSB activity that happens *before* significant changes in the stock market as effective or at least suggestive of WSB being a good source of information, even if the primary cause lies somewhere else and WSB only picked up on the development.

### Reliability of Buy and Sell Signals

The average performance of the portfolio has been good, but has individual investment advice from posts on r/WallStreetBets been reliable? For each of the WSB portfolio's tickers, we evaluated the accuracy of the community's buy and sell signals by detecting all signals and comparing them with the respective price movements over specific time windows – if the price increased after a buy signal, it is considered successful (similarly with price decreases for sell signals).

The first comparison is intended to show if the buy and sell signals would have led to better investment decisions than investing continuously or randomly in the same stocks over the same time frame, with a focus on short-term windows. To this end, the success rate of all buy signals for a given stock is compared to how successful an investment would have been if bought or sold

**a.** equally distributed – i.e., bought or sold on equally distributed days (same number of days as respective original signal) based on a regular time interval,

**b.** randomly distributed – i.e., bought or sold on randomly chosen days (same number of days as in respective original signal, but computed as an average over 5 independent trials for a more robust estimate),

**c.** every day – i.e., bought or sold on every single day possible.

More formally, let $X_i \in \mathcal{X}$ denote the $i^{\text{th}}$ data point from the reviewed dataset $\mathcal{X}$ for $i \in \{1, \ldots, D\}$, where $D$ is the total number of days in dataset. Each such $X_i$ provides the pertinent data for a stock at day $i$, including the stock value and its relative price changes. Further let $n$ denote the total number of a signal type (buy or sell) within the corresponding time period. The three different baselines are then defined as follows:

**a.** Equally distributed sample of investment days using the same sample size as the actual amount of buy signals detected per ticker: We consider all samples $X_i \in \mathcal{X}$ such that
  - $i = \lfloor \frac{D}{n} \rfloor k + \delta$ for $k \in \{1, \ldots, n\}$ and $\lfloor \frac{D}{n} \rfloor k + \delta \leq D$,
  - where $\delta$ is an optional offset to avoid all $X_i$ falling on the weekend. It is $0$ in the vast majority of cases, but instead set to $\lfloor \frac{1}{2} \frac{D}{n} \rfloor$ in the special circumstance of $\lfloor \frac{D}{n} \rfloor = 7$ and simultaneously all $X_i$ for $i = \lfloor \frac{D}{n} \rfloor k$ falling on the weekend.

**b.** Randomly distributed values: We select $n$ samples of $X_i$ such that
  - $i \in I$ where $I$ is a set of $n$ random numbers from $\{1, \ldots, D\}$ sampled without replacement,
  - and, in order to increase robustness, the procedure is repeated across 5 trials and the average is computed.

**c.** Every day: We consider
  - all $X_i$ within the date range, i.e., for all $i \in \{1, \ldots, D\}$.

Table 5 reports the average price change after predefined time windows depending on when an investment was made. If an investment in a WSB portfolio stock was made following a buy signal, the price increased after one day in 51.75% of the cases, after three days in 52.88%, after one week in 55.02%, after one month in 61.28%, and after three months in 69.94% of cases.

The baselines simulate three alternative types of hypothetical investors: one that invests every $\lfloor \frac{D}{n} \rfloor$ days, one that chooses $n$ random days to invest, and one that invests every day (with $n$ being equal to the number of buy or sell signals that the baseline is compared to and $D$ being the total number of days in the dataset). The results show that between the three different baseline distributions, the differences are not particularly strong. A comparison of the buy signals' success rates show that within the shorter time frames of one day to one week, following the buy signals is approximately equally often successful as buying on equally or randomly distributed days or on an average day. However, following the buy signals seems to be more successful across longer time frames, as the buy signals' success rate is up to 5.4% higher (after one month) and up to 15.6% higher (after three months) compared to the baselines' success rates. This means that when an investment decision was made based on a buy signal, in almost 70% of the cases the price increased within the following three months. If, in contrast, investments were made based on equally or randomly distributed transactions, the price increased within the following three months in only 60–62% of the cases.

| | Success rate of (hypothetical) investment after | | | | |
| | 1 day | 3 days | 1 week | 1 month | 3 months |
|---|---|---|---|---|---|
| buy signals | 51.75% | 52.88% | 55.02% | 61.28% | 69.94% |
| equally distr. | 52.80% | 53.19% | 54.40% | 59.19% | 61.28% |
| randomly distr. | 51.51% | 53.12% | 54.68% | 58.12% | 60.52% |
| every day | 51.08% | 52.70% | 54.93% | 59.12% | 61.60% |

**Table 5. Average success rate of buy signals (price change > 0, average of all tickers in WSB portfolio) compared to equally and randomly distributed signals, and ratio of all positive price developments**

The results from Table 5 imply that with regard to a true / false classification, WSB's buy signals do not perform much better than random or equally distributed signals in the short term – a difference only becomes discernible when looking at time windows of one month or longer. However, comparing general upward or downward movement of prices after investment only provides a very coarse-grained assessment.

In the next step, a review of the average price changes yields more detailed insights. As shown in Table 4, making an investment on days of ticker mentions or buy signals would have led to significantly higher growth than distributing investments similarly to the average of all days within the range. Table 6 adds price development values extracted from the equally and randomly distributed baselines described above – and the values suggest that investment decisions based on WSB activity (mentions and buy/sell signals) entailed larger subsequent price increases. Again, it should be noted that sell signals have been highly unsuccessful, as the prices increased even more after those days than when selling on days without community activity.

| | t | Total Avg. | When bought on specific days: | | | | |
| | | | Mention | Buy Sig. | Sell Sig. | Eq. Distr. | Rnd. Distr. |
|---|---|---|---|---|---|---|---|
| Avg. price change (%) after | 1 day | 0.17 | 0.29 | 0.29 | 0.65 | 0.19 | 0.13 |
| | 1 week | 1.24 | 2.07 | 2.19 | 3.77 | 1.21 | 1.44 |
| | 1 month | 5.14 | 7.77 | 9.88 | 8.12 | 5.10 | 5.36 |
| | 3 months | 16.65 | 27.32 | 32.03 | 28.53 | 14.64 | 15.86 |

**Table 6. Average WSB portfolio development for different investment patterns: based on ticker mentions, buy signal, sell signal, equal distribution, random distribution, all (January 2019 – April 2021)**

For a more fine-granular analysis, the comparison of the price development of buy signals versus the average may be reviewed on a per-stock basis. Table 7 reports the price development for the WSB portfolio stocks (except $SPY) for investments following a buy signal as well as the average price development. This detailed overview shows that following buy signals has only led to a consistently better outcome for a few selected stocks: $GME, $AMC, $BB, $AMZN, $GM, $TLRY, $TSLA – the first three ($GME, $AMC, $BB) belong to the four *meme stocks* that were at the center of attention at the end of January 2021, $TLRY is a cannabis stock that achieved *meme stock* status in February due to its sudden increase in popularity, and $TSLA has been a highly popular stock throughout the last years. While $GM has not shown *meme stock* characteristics (extremely high popularity in the short term coupled with high price spike and volatility), it appears to be a stock that is often mentioned and discussed in the WSB community. However, WSB did not produce reliable buy signals for the other prominent *meme stocks* $NOK and $ACB.

When only comparing price developments after three months, a few tickers can be added to the list above: $AAPL, $AMZN, $BA, $FB, $GE, $NIO, $QQQ – following buy signals with these stocks resulted in higher growth than the average price development. This means that despite volatility in the short term, following buy signals would have turned profitable after three months for 14 of the 20 tickers.[9] When excluding

---

[9] Arguably, we could also include $NOK in this list, as the loss of an investor following buy signals would have been less than the average

the stocks that have incurred losses on average ($NOK, $ACB, $BA), following buy signals achieved a 56% (mean) or 17% (median) higher price growth than the average price increase.

In conclusion, we need to distinguish two cases:

- in the short term (1 day, 1 week) an investment strategy following buy signals actually has not been more successful than the average price increase with respect to short-term windows since January 2019 – instead the success of buy signals was mostly driven by *meme stocks*, which biased the average results (especially GameStop),
- in the longer term (3 months), however, the buy signals turned out to provide better growth than the average price development on average.

This confirms the findings of the previous analysis: trusting buy signals on the short term seems to be successful as often as it fails, but in principle, the portfolio of WSB's preferred stocks has performed well and the buy signals have turned out to be valuable indicators for more patient investors.

## Proactive vs. Reactive Signals

The previous analysis explains that while WSB has shown a tendency to bet on stocks that were likely to perform better than the broader market, an investment strategy following buy signals would not have been much more successful than the average price development in the short term. With the goal of separating valuable signals from those that are unsuccessful, the following analysis investigates whether signals can be classified as proactive or reactive. The motivation for this distinction is that the two can have very different implications regarding their reliability if taken as investment advice. Reactive signals (e.g., a buy signal just after the stock price rose significantly or a sell signal just after a significant price drop) are usually not helpful in the short term – on the contrary, they can quickly lead to financial losses if taken as investment advice, e.g., shortly after a significant price increase in a volatile market. However, they occur very frequently due to users sharing their gains or losses from recent investments or posting news articles on specific stocks that recently made headlines due to stronger price changes. The most valuable form of short-term investment advice is one received before a change in stock price occurs, thus constituting a proactive signal.

We consider a simple definition for each case (with $x$ days referring to 1 day, 3 days, or 1 week):

- **Reactive:** the price increase in the preceding $x$ days is *higher* than the price change in the following $x$ days, e.g., price movement from $10 on May 1 to $17 on May 2 and then to $16 on May 3.
- **Proactive:** the price change in the preceding $x$ days is *lower* than the price increase in the following $x$ days, e.g., price movement from $10 on May 1 to $11 on May 2 and then to $20 on May 3.

Following this classification, only 46.5% of all buy signals were proactive on average, while slightly more than half were reactive and thus less successful. This suggests that a separation of signals into proactive and reactive is important for extracting the most valuable pieces of investment advice. It should be noted that we do not apply this analysis to sell signals, as the previous section has shown that sell signals have been quite unsuccessful in the reviewed time frame. Table 8 provides a detailed overview of how the prices developed before and after buy signals, distinguishing between baseline (average of all days as well as days with buy signals only) versus reactive and proactive signals, as detected using three different time windows $x$.[10]

Table 8 reveals clearly different patterns for reactive and proactive signals: Investments following reactive signals performed significantly worse than those following proactive signals – indicating that the price performance of all buy signals seem to be an average of these two different types of signals. Nonetheless, even the reactive investments achieved higher growth than the average price increase. The effect is most pronounced for a time period of one week.

Reactive and proactive buy signals can be identified by comparing the average performance since the previous week with the average performance of all buy signals – if the price increase since the previous week was

---

price decrease. However, as the investment would have resulted in a loss, this case is excluded in the evaluation.
[10]Three different $x$ are used: 1 day, 3 days, 1 week. $y$ additionally includes 1 and 3 months.

| Ticker | Pattern | Avg. price change (%) after | | | |
|--------|---------|-------|--------|---------|----------|
| | | 1 day | 1 week | 1 month | 3 months |
| GME | Average | 0.76 | 7.00 | 29.95 | 109.16 |
| | Buy Signal | 2.86 | 22.35 | 100.66 | 355.36 |
| AMC | Average | 0.40 | 2.30 | 8.70 | 8.80 |
| | Buy Signal | 0.92 | 11.50 | 33.75 | 28.11 |
| BB | Average | 0.10 | 0.73 | 3.68 | 10.43 |
| | Buy Signal | 0.10 | 1.12 | 4.54 | 14.38 |
| NOK | Average | 0.00 | -0.06 | -0.52 | -2.29 |
| | Buy Signal | -0.28 | -0.64 | -3.00 | -0.69 |
| AAPL | Average | 0.16 | 1.12 | 4.98 | 16.08 |
| | Buy Signal | 0.14 | 0.95 | 4.36 | 16.39 |
| ACB | Average | -0.04 | -0.24 | -2.73 | -10.25 |
| | Buy Signal | 0.29 | -1.39 | -4.45 | -9.38 |
| AMD | Average | 0.22 | 1.48 | 6.19 | 19.61 |
| | Buy Signal | 0.09 | 1.35 | 5.34 | 17.10 |
| AMZN | Average | 0.10 | 0.68 | 2.74 | 9.12 |
| | Buy Signal | -0.01 | 0.77 | 3.19 | 11.80 |
| BA | Average | 0.03 | 0.34 | 0.80 | -1.56 |
| | Buy Signal | 0.08 | 0.03 | -0.58 | 0.46 |
| BABA | Average | 0.08 | 0.54 | 2.11 | 5.97 |
| | Buy Signal | 0.12 | 0.23 | 1.22 | 4.18 |
| FB | Average | 0.12 | 0.78 | 3.03 | 7.81 |
| | Buy Signal | 0.20 | 0.69 | 2.89 | 9.56 |
| GE | Average | 0.11 | 0.72 | 2.95 | 7.56 |
| | Buy Signal | -0.09 | 0.44 | 2.54 | 10.17 |
| GM | Average | 0.10 | 0.74 | 2.86 | 7.72 |
| | Buy Signal | 0.15 | 1.27 | 5.47 | 15.90 |
| INTC | Average | 0.07 | 0.45 | 1.84 | 3.83 |
| | Buy Signal | 0.20 | 0.10 | 1.57 | 2.54 |
| MSFT | Average | 0.12 | 0.82 | 3.44 | 10.42 |
| | Buy Signal | 0.10 | 0.70 | 2.44 | 10.11 |
| NIO | Average | 0.36 | 2.76 | 13.60 | 62.35 |
| | Buy Signal | 0.02 | 0.36 | 12.30 | 69.92 |
| NVDA | Average | 0.21 | 1.38 | 5.53 | 18.10 |
| | Buy Signal | 0.10 | 1.44 | 3.75 | 17.03 |
| QQQ | Average | 0.10 | 0.69 | 2.85 | 8.74 |
| | Buy Signal | 0.03 | 0.35 | 2.20 | 11.39 |
| TLRY | Average | 0.08 | 0.80 | 2.02 | 7.54 |
| | Buy Signal | 0.46 | 0.45 | 14.12 | 35.62 |
| TSLA | Average | 0.35 | 2.54 | 11.95 | 45.19 |
| | Buy Signal | 0.55 | 3.49 | 13.31 | 47.18 |

**Table 7. Average price change vs. average price change after buy signal for selected WSB portfolio stocks (January 2019 – April 2021)**

| | t | Baseline | | w/ Reactive Buy Sig. | | | w/ Proactive Buy Sig. | | |
|---|---|---|---|---|---|---|---|---|---|
| | | Avg | Buy Sig. | $x = 1d$ | $x = 3d$ | $x = 1w$ | $x = 1d$ | $x = 3d$ | $x = 1w$ |
| AVG perf (%) since | 1w | 1.24 | 4.15 | 8.96 | 11.88 | 14.55 | 2.63 | -0.07 | -1.33 |
| | 3d | 0.51 | 1.76 | 6.53 | 8.45 | 6.18 | 0.44 | -1.37 | -0.21 |
| | 1d | 0.17 | 0.52 | 4.85 | 2.98 | 2.08 | -1.23 | -0.67 | -0.21 |
| AVG perf (%) after | 1d | 0.17 | 0.29 | -1.24 | -0.39 | -0.23 | 4.57 | 2.30 | 1.76 |
| | 3d | 0.51 | 0.90 | -0.67 | -1.64 | -1.31 | 5.13 | 7.68 | 5.85 |
| | 1w | 1.24 | 2.18 | 1.17 | -0.93 | -3.15 | 6.40 | 10.11 | 13.64 |
| | 1m | 5.14 | 9.88 | 6.55 | 4.20 | 2.00 | 11.85 | 14.89 | 18.13 |
| | 3m | 16.65 | 32.03 | 25.77 | 21.89 | 21.09 | 37.86 | 44.05 | 51.35 |

**Table 8. Comparison of reactive vs. proactive signals as well as baselines (average and all buy signals) with regard to average price performance over preceding $x$ days as well as average price development after $y$ days (WSB portfolio, January 2019 – April 2021)**

2–3 times higher than the average after a buy signal and the price has already increased by more than 5% since the previous day, chances are high that the day's buy signal is reactive.

Thus, reactive signals can be detected based on a substantially above average price development in the time before the buy signal, while the price performance before a proactive buy signal is below average with negative price movement since the previous day on average. By applying this technique, an investor following only proactive buy signals would have been able to achieve up to 700% higher growth after a single day and up to 50% higher growth after three months on average. This conclusion also applies when excluding the *meme stocks* of January 2021 from the calculation – in this case, up to 2,400% higher growth after a single day and up to 75% higher growth after three months could be achieved on average. This shows that once being able to distinguish reactive from proactive buy signals, investors relying on the WSB community for investment decisions can not only increase their longer term investment success, but also perform successful short term investments.

A stock-specific review of the values given in Table 8 confirms these results: investments made after reactive signals follow a period of above average price increase, provide a below-average (compared to all buy signals) return after investment, and show higher volatility and trading volume on days of the signal. In contrast, proactive buy signals follow a period of below-average price movement (often a price decrease since the last few days), provide an above-average growth, and tend to coincide with below-average volatility and trading volume. These effects are amplified when the time $x$ for detecting a reactive or proactive signal is longer (one week) rather than only a day, but apply for all evaluated $x$ values.

Clearly, as a stock's future price development is unclear at the time of posting, it is much easier to decide retrospectively whether a signal was proactive or reactive. Recognizing proactive signals directly at the time of posting is more challenging, as future price developments are uncertain. Nonetheless, we have identified a simple heuristic for making this distinction using historic price information only: We compare the price development for the preceding one week, three days, and one day with their respective moving averages over 30 days. This methodology succeeds at identifying proactive buy signals with regard to their price development after one month and three months: If any of the values were below their respective moving averages, an investment following the day's buy signal would have achieved a more successful performance than trusting all buy signals on average – buy signals on days with a price change in the preceding one day that is below the 30-day moving average achieved approx. 83% higher growth after one day and approx. 17% higher growth after three months than the respective values of all buy signals on average. If all three values were below their respective moving averages, the average price growth improved as well, with 25% higher growth after one day and approx. 44% higher price increase. While the 30-day moving averages did not enable proactive-signal performance on the short term of one day to one week, the growth after one and three months are very close to the average of proactive signals. This shows that even with a simple technique to assess the validity of buy signals, investments in the WSB portfolio would have been even more profitable

| | | Jan. 2019 – Apr. 2021 | | Jan. 2019 – Dec. 2020 | |
|---|---|---|---|---|---|
| | t | All | w/ Buy Sig. | All | w/ Buy Sig. |
| Avg. price change (%) after | 1 day | 0.17 | 0.29 | 0.11 | 0.11 |
| | 1 week | 1.24 | 2.19 | 0.78 | 0.49 |
| | 1 month | 5.14 | 9.88 | 3.15 | 3.60 |
| | 3 months | 16.65 | 32.09 | 8.72 | 12.57 |
| Average | mentions | 28.28 | 89.50 | 7.70 | 14.90 |
| | daily volat. | 2.9% | 4.2% | 2.7% | 3.7% |
| | daily volume | 25.56M | 37.95M | 24.36M | 33.10M |

**Table 9. Comparison of full vs. pre-hype phases with regard to average price change per time window, mentions, daily volatility, and daily volume (for all days or only days with buy signals)**

| | | Jan. 2019 – Apr. 2021 | | | | | Jan. 2019 – Dec. 2020 | | | | |
|---|---|---|---|---|---|---|---|---|---|---|---|
| | | All Buy | React. Buy | | Proact. Buy | | All Buy | React. Buy | | Proact. Buy | |
| | t | | 1d | 1w | 1d | 1w | | 1d | 1w | 1d | 1w |
| % chg. since | 1w | 4.15 | 8.96 | 14.55 | 2.63 | -1.33 | 1.91 | 5.76 | 9.70 | 0.47 | -2.86 |
| | 3d | 1.76 | 6.53 | 6.18 | 0.44 | -0.21 | 0.78 | 4.56 | 4.42 | -0.50 | -1.19 |
| | 1d | 0.52 | 4.85 | 2.08 | -1.23 | -0.21 | 0.18 | 3.90 | 1.61 | -1.40 | -0.77 |
| % chg. after | 1d | 0.29 | -1.24 | -0.23 | 4.57 | 1.76 | 0.11 | -1.05 | -0.42 | 3.84 | 1.70 |
| | 3d | 0.90 | -0.67 | -1.31 | 5.13 | 5.85 | 0.24 | -1.03 | -1.34 | 3.79 | 4.65 |
| | 1w | 2.18 | 1.17 | -3.15 | 6.40 | 13.64 | 0.49 | -0.75 | -2.92 | 4.29 | 9.51 |
| | 1m | 9.88 | 6.55 | 2.00 | 11.85 | 18.13 | 3.60 | 1.56 | -0.08 | 8.64 | 13.65 |
| | 3m | 32.03 | 25.77 | 21.09 | 37.86 | 51.35 | 12.57 | 10.64 | 8.04 | 17.12 | 25.19 |

**Table 10. Comparison of full vs. pre-hype phases with respect to average price change of WSB portfolio in preceding $x$ days and subsequent $y$ days following all buy signals (average), only reactive buy signals, and only proactive signals.**

with regard to the longer time windows.

## Before and After the Hype

It is important to note that the rapid increase in community members and generated content in and after January 2021 entails a strong shift in the dataset towards the newest posts and discussions, which were also coupled with high-volatility price movement in the financial markets. In order to assess whether the observations from the analysis above also hold for the time before the January hype phase, we applied the same analysis to the dataset limited to the window from January 1, 2019 to December 31, 2020.

Table 9 shows that for the time frame before 2021, buy signals were much less effective over the short term than in 2021, but still showed better performance over longer time periods than trusting all buy signals. The overall impact of buy signals is weaker in contrast to the post-hype values – nonetheless, similar tendencies can be detected. We obtain the same for the comparison of reactive and proactive signals – while the values are not as extreme, the tendencies and therefore conclusions remain the same. Table 10 presents a brief comparison of proactive and reactive signals showing these less pronounced yet similar tendencies.

# Conclusion

Reddit's r/WallStreetBets has come to prominence as a forum for unconventional high-risk investment discussions. This paper assesses the reliability of investment advice encountered in the subreddit. The analysis reveals that recent activity on r/WallStreetBets (since January 2019) could have served as profitable investment advice on multiple levels:

1. WSB's discussions have been focused mostly on sectors that have performed better over the last three years and last year than the broader market referenced by the S&P500 index. Investing in these sectors and holding for a longer period would have yielded better results than the average market growth.
2. A hypothetical portfolio containing the most frequently and consistently discussed stocks has outperformed the S&P500 on average over the last three years as well as the last year.
3. An investment strategy that followed buy signals would have led to more successful average growth on the longer term than distributing investments equally or randomly over the same time window – while in the short term the results of these different strategies would have been similar.
4. Investments following only proactive buy signals and ignoring reactive signals would have achieved even more successful outcomes on average and also achieved short term success in investments.
5. The tendencies in the results are amplified by WSB's strong growth and the success of *meme stocks* in 2021, but they can still be found in a weaker form when only analysing data from January 2019 to December 2020, thus corroborating the above analysis results.